\documentclass[]{spie}  

\usepackage{amsmath,amsfonts,amssymb}
\usepackage{graphicx}
\usepackage[normalem]{ulem}
\usepackage[colorlinks=true, allcolors=blue]{hyperref}

\let\citep\cite
\let\citet\cite

\graphicspath{{paper_figures/}}

\title{Instrument response generation using high-resolution 3D voxelization of GRBAlpha and VZLUSAT-2 satellites with MEGAlib}

\author[a*]{Jean-Paul Breuer}
\author[a\textdagger]{Masato Yokota}
\author[a]{Yasushi Fukazawa}
\author[a]{Hiromitsu Takahashi}
\author[b]{Norbert Werner}
\author[b]{Jakub \v{R}\'{i}pa}
\author[b]{Marianna Daf\v{c}\'{i}kov\'{a}}
\author[b]{Filip Munz}
\author[a]{Masanori Ohno}

\author[c]{Bal\'{a}zs Cs\'{a}k}
\author[c]{L\'{a}szl\'{o} M\'{e}sz\'{a}ros}
\author[c]{Andr\'{a}s P\'{a}l}

\author[d]{Marcel Frajt}
\author[d]{Jan Hudec}
\author[d]{Jakub Kapu\v{s}}
\author[d]{Maksim Rezenov}

\author[e]{Vladim\'{i}r D\'{a}niel}
\author[e]{Juraj Dud\'{a}\v{s}}
\author[e]{Petr Svoboda}

\author[f,g]{Hsiang-Kuang Chang}
\author[g]{Hao-Min Chang}
\author[h]{Chin-Ping Hu}
\author[i]{Chih-Hsun Lin}
\author[j]{Tsung-Che Liu}
\author[f]{Kaustubha Sen}
\author[k]{Che-Chih Tsao}
\author[f]{Chih-En Wu}

\affil[a]{Department of Physics, Graduate School of Advanced Science and Engineering, Hiroshima University Kagamiyama, 1-3-1 Higashi-Hiroshima, 739-8526, Japan}
\affil[b]{Department of Theoretical Physics and Astrophysics, Faculty of Science, Masaryk University, Brno, Czech Republic}

\affil[c]{Konkoly Observatory, Research Centre for Astronomy and Earth Sciences, Budapest, Hungary}
\affil[d]{Spacemanic Ltd., Bratislava, Slovakia}
\affil[e]{Czech Aerospace Research Centre, Prague, Czech Republic}
\affil[f]{Institute of Astronomy, National Tsing Hua University, Hsinchu, Taiwan, Republic of China}
\affil[g]{Department of Physics, National Tsing Hua University, Hsinchu, Taiwan, Republic of China}
\affil[h]{Department of Physics, National Changhua University of Education, Changhua City, Taiwan, Republic of China}
\affil[i]{Institute of Physics, Academia Sinica, Taipei, Taiwan, Republic of China}
\affil[j]{Department of Applied Physics, National Pingtung University of Science and Technology, Taiwan, Republic of China}
\affil[k]{Department of Power Mechanical Engineering, National Tsing Hua University, Hsinchu, Taiwan, Republic of China}

\authorinfo{* E-mail: jeanpaul.breuer@gmail.com; \textdagger \,E-mail: masatotb9796@gmail.com}

\begin{document} 
\maketitle

\begin{abstract}
Pathfinder gamma-ray burst (GRB) detecting CubeSat missions such as GRBAlpha and VZLUSAT-2 have demonstrated the successful application of scintillator detectors with silicon photomultipliers in low Earth orbit (LEO). To produce more comprehensive scientific analysis of the data, the effective area of the detector needs to be characterised at different energies. A large part of this process requires a thorough understanding of the detectors response matrices based on the satellite mass model typically performed through Geant4 and MEGAlib simulations. We use a novel voxelization and binning methodology to turn complex 3D geometries into MEGAlib-compatible versions, and we validate these experiments by showing that the simulation results with Geant4 agree within an order of 10\%. We use MEGAlib simulations from various angles around the spacecraft to generate instrumental response matrices for both simple and complex geometries, in the case of GRBAlpha and VZLUSAT-2, respectively. 
\end{abstract}


\section{Introduction}
\label{sec:intro}
Gamma-ray bursts (GRBs) are among the most energetic transient phenomena in the Universe, releasing a large amount of energy on timescales from milliseconds to hundreds of seconds. Because their occurrence time and sky position cannot be predicted in advance, GRB observations fundamentally require wide-field, nearly continuous monitoring. This requirement has become even more important in the era of multi-messenger astronomy, where short GRBs are regarded as key electromagnetic counterparts to compact-object mergers and rapid localization is needed to enable follow-up observations at other wavelengths and with other messengers.

A single observatory, however, faces an inherent trade-off between field of view, sensitivity, and localization accuracy. Constellation-based observations with multiple small satellites provide a practical way to overcome this limitation by combining all-sky coverage with localization based on inter-satellite arrival-time differences and detector response anisotropy. Within this context, the CAMELOT mission aims to use a network of CubeSats carrying CsI(Tl)+SiPM gamma-ray detectors to detect and localize GRBs with substantially improved sky coverage and localization capability. GRBAlpha and VZLUSAT-2 serve as important pathfinder missions in this program, demonstrating that compact CubeSat platforms can routinely detect high-energy transients in low Earth orbit and can provide the experimental basis for future constellation observations.

To interpret the observational data quantitatively, accurate detector response matrices (DRMs) are required for each satellite and for each incident direction. In small satellites, the detector response depends not only on the scintillator itself, but also on surrounding spacecraft structures such as shields, batteries, printed circuit boards, solar panels, and support frames. This makes detailed mass-model simulations essential. Geant4 offers high flexibility for such simulations, especially when CAD-based geometries are available, whereas MEGAlib provides a powerful end-to-end framework for gamma-ray detector simulations and response generation. The difficulty is that MEGAlib geometry descriptions are best suited to primitive solids, so complex spacecraft components must be approximated efficiently without losing the shielding and scattering properties that dominate the response.

In this work, we develop a practical workflow for response generation that bridges this gap. We convert CAD-derived spacecraft components into high-resolution voxelized models, compress them using a greedy brick-merging procedure, and export them as MEGAlib-compatible geometries. We then perform mono-energetic simulations over multiple incident directions to construct DRMs and effective-area curves for GRBAlpha and VZLUSAT-2. For GRBAlpha, the MEGAlib-derived response is validated against an existing Geant4 reference response. For VZLUSAT-2, we apply the same workflow to a more complex 3U geometry and compare simplified and voxel-optimized models to evaluate the trade-off between response fidelity and computational cost.

Section 2 introduces the GRBAlpha and VZLUSAT-2 satellites in more detail. Section 3 discusses the main methodology for preparing MEGAlib geometry, as well as the experimental framework in Geant4 and MEGAlib for the response generation. Section 4 validates the simulation results and compares the response products obtained with Geant4 and MEGAlib. Section 5 concludes with the main findings and future activities.

\section{GRBAlpha and VZLUSAT-2 CubeSats}\label{sec:data} 
GRBAlpha \citep{Pal2021,Pal2023} was a 1U CubeSat which launched on March 22, 2021 at 550 km altitude in a sun-synchronous polar orbit (SSO). The smallest astrophysical space observatory, GRBAlpha operated as a proof-of-concept for a future constellation of nanosatellites, CAMELOT \citep{CAMELOT}. The detector consisted of a 75 $\times$ 75 $\times$ 5 mm CsI(Tl) scintillator read out by eight multi-pixel photon counters (MPPCs) S13360-3050 PE produced by Hamamatsu Photonics K.K., arranged into two independent readout channels with four MPPCs connected in parallel in each readout channel \citep{Ripa2025}. With an effective energy range between 30 - 900 keV at the start of the mission, GRBAlpha successfully detected a total of 227 transients, consisting of 124 GRBs (20 short GRBs and 104 long GRBs), 100 solar flares, 2 magnetar-related events, and 1 X-ray binary outburst\footnote{\href{https://monoceros.physics.muni.cz/hea/GRBAlpha/}{https://monoceros.physics.muni.cz/hea/GRBAlpha/}}, during its lifetime before its eventual re-entry into Earth's atmosphere on June 9, 2025 \citep{Pal2025,grbstats}.

VZLUSAT-2 \citep{Daniel2020} is a 3U CubeSat mission developed by the Czech Aerospace Research Center (VZLU), which was launched on January 13, 2022 to a 535 km altitude SSO. The primary payloads on VZLUSAT-2 are two Earth-observing cameras, but additionally include two perpendicularly placed GRB Detectors, a solar X-ray monitor, and radiation dosimeters as secondary instruments. Specifically, the two GRB detectors are almost identical to the one on GRBAlpha. Besides some minor modifications to the mounting, the key differences compared to GRBAlpha were in how the analog frontend electronics boards are housed, and with how the MPPCs are coupled to the scintillator \citep{Ripa2025}. VZLUSAT-2 re-entered Earth's atmosphere on November 30, 2025 and over approximately four years of operation, detected a total of 146 transients, consisting of 68 GRBs (8 short GRBs and 60 long GRBs), 73 solar flares, and 5 magnetar-related events\footnote{\href{https://monoceros.physics.muni.cz/hea/VZLUSAT-2/}{https://monoceros.physics.muni.cz/hea/VZLUSAT-2/}} \citep{grbstats}.

Together, GRBAlpha and VZLUSAT-2 have both successfully demonstrated that CubeSats can be routinely used to monitor gamma-ray transients in the sky, at a fraction of the cost of traditional large-class missions.

\section{Methods}
\label{sec:methods}
\subsection{CAD geometry to MEGAlib conversion workflow}
\begin{figure}[h]
\centering
\includegraphics[width=0.96\linewidth]{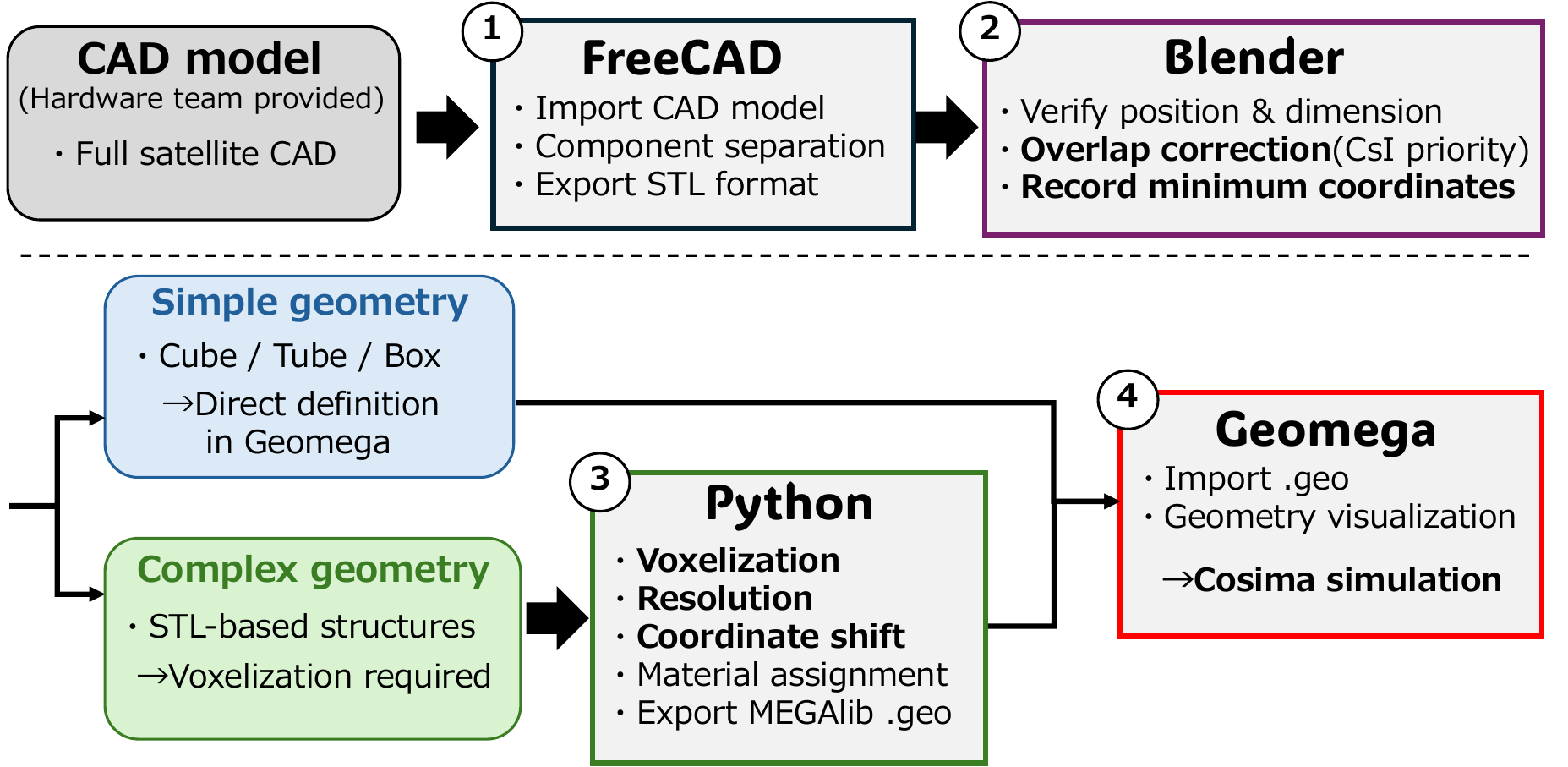}
\caption{CAD-to-MEGAlib conversion workflow used in this work. Simple components are described directly with primitive solids, while complex structures are voxelized and then exported as MEGAlib geometry files.}
\label{fig:pipeline}
\end{figure}

The CAD to MEGAlib geometry workflow can be effectively summarized in Figure~\ref{fig:pipeline}, and involves several stages. Beginning with the full CAD model, critical components are first separated and individually exported as STL geometry files. These STL files are each imported into Blender, before their coordinates, rotation, and dimensions, are recorded and verified. In the case of overlapping components, a manual correction must be performed to remove the overlap, before being again exported as ascii-type STL files if they were not already in the ascii format. Finally, dependent on geometry complexity, they are either manually defined with simple MEGAlib geometry, or voxelized through the Python routine described in detailed below.

All of the relevant code described below is open source and available on GitHub at the STL\_to\_MEGAlib repository\footnote{\href{https://github.com/jpbreuer/STL_to_MEGAlib}{https://github.com/jpbreuer/STL\_to\_MEGAlib}}
\subsubsection{3D model voxelization}
To fully utilize the power of MEGAlib, it is necessary to create the closest, most realistic approximation of a mass model as possible to the original spacecraft and detector geometry, including the density and composition of every component on the spacecraft. To do this, we first export each component of the 3D model geometry as individual STL files, then use a voxelization procedure to convert the triangular STL mesh into a three-dimensional matrix with either a user-specified resolution along the $z$ direction, or a given voxel size in physical units. The overall conversion flow from hardware CAD to simulation-ready MEGAlib geometry is summarized in Figure~\ref{fig:pipeline}. The procedure was based on the stl-to-voxel code of
Christian Pederkoff\footnote{\href{https://github.com/cpederkoff/stl-to-voxel}{https://github.com/cpederkoff/stl-to-voxel}} which additionally features an updated fast winding number based algorithm for polygon repair \citep{Segmentation2013}. A polygon repair method is necessary to guarantee watertightness at the voxel level, since STL models may produce fragmented polylines and spurious gaps caused by floating-point edge cases and numerical coincidences at the slice height.

We keep voxel edges $(\Delta_x,\Delta_y,\Delta_z)$ uniform by construction; when a single physical size $\Delta$ is provided, $\Delta_x=\Delta_y=\Delta_z=\Delta$. The geometric error of voxelization is bounded by half the voxel diagonal,
\[
\varepsilon_{\mathrm{vox}} \le \tfrac{1}{2}\sqrt{\Delta_x^2+\Delta_y^2+\Delta_z^2},
\]
while STL model geometric properties (volume, position) are conserved up to this discretization. Features smaller than one voxel vanish, so users should increase resolution in such cases, at the expense of upstream memory and computation time during simulation. To interface with MEGAlib (which uses centimeters), we convert all lengths to cm when exporting.

\subsubsection{Greedy binning}
\label{sec:greedy-binning}
Choosing a high-resolution description can result in potentially hundreds of millions of tiny voxels for a single object. To comply with the MEGAlib geometry description requirement, each individual voxel must be defined by unique identifier, as well as an assigned shape and material composition. As such, without any binning strategy, can result in output MEGAlib geometry files on scales of a few kilobytes to the order of several Gigabytes, depending on the chosen resolution. Importing Gigabyte size geometry files dramatically increases the computational and memory requirements needed to perform simulations, so we implemented a binning algorithm as a way to minimize the number of total individual voxels requires to describe a given 3D object.  

\paragraph{Greedy maximal-extent decomposition.}
The greedy binning method compresses the voxelized solid into a small set of disjoint, axis-aligned rectangular boxes whose union equals the filled set exactly; in other words, the method merges adjacent voxels together into larger voxels, and never revisits a previous voxel. Our \emph{voxel-to-bricks} routine performs a single pass that constructs \emph{maximal} boxes with an $x\!\rightarrow\!y\!\rightarrow\!z$ expansion order:
\begin{enumerate}
\item Scan each slice row-by-row. When an unvisited filled voxel at $(x,y,z)$ is found, greedily extend a run $[x_0,x_1]$ along $+x$ while voxels remain filled and unvisited.
\item Attempt to grow along $+y$: accept a new row $y_1{+}1$ iff \emph{every} voxel in $[x_0,x_1]$ is filled and unvisited at that row. Repeat until failure.
\item Attempt to grow along $+z$ with the same all-voxels test applied to the current $(x,y)$ rectangle across slices.
\item Mark the resulting block $[x_0,x_1]\times[y_0,y_1]\times[z_0,z_1]$ as visited and append it to the brick list.
\end{enumerate}
Because each voxel is visited once and claimed by exactly one brick, the decomposition is exact and the bricks are pairwise disjoint. Each brick is \emph{maximal} with respect to the chosen axis order; it cannot be enlarged by one voxel along any axis without crossing empty or already-claimed space.

Because each voxel is visited at most once, the routine runs in $O(N_xN_yN_z)$ time and has worst-case $O(N_xN_yN_z)$ memory requirements, where $N_x$, $N_y$, and $N_z$ are the number of voxels along the three grid axes. For ``compact'' solids with large uniform regions, the brick count collapses dramatically, from tens of millions of voxels to a few hundred bricks in our tests. In adversarial cases (e.g., a 3D checkerboard) limited merging is possible. While a globally minimum brick cover is computationally hard in general, the greedy strategy provides an excellent size–speed trade-off and deterministic runtime—appropriate \cite{Temlyakov2003,Theodoridis2015}.

\begin{figure}[h]
\centering
\includegraphics[width=0.666\linewidth]{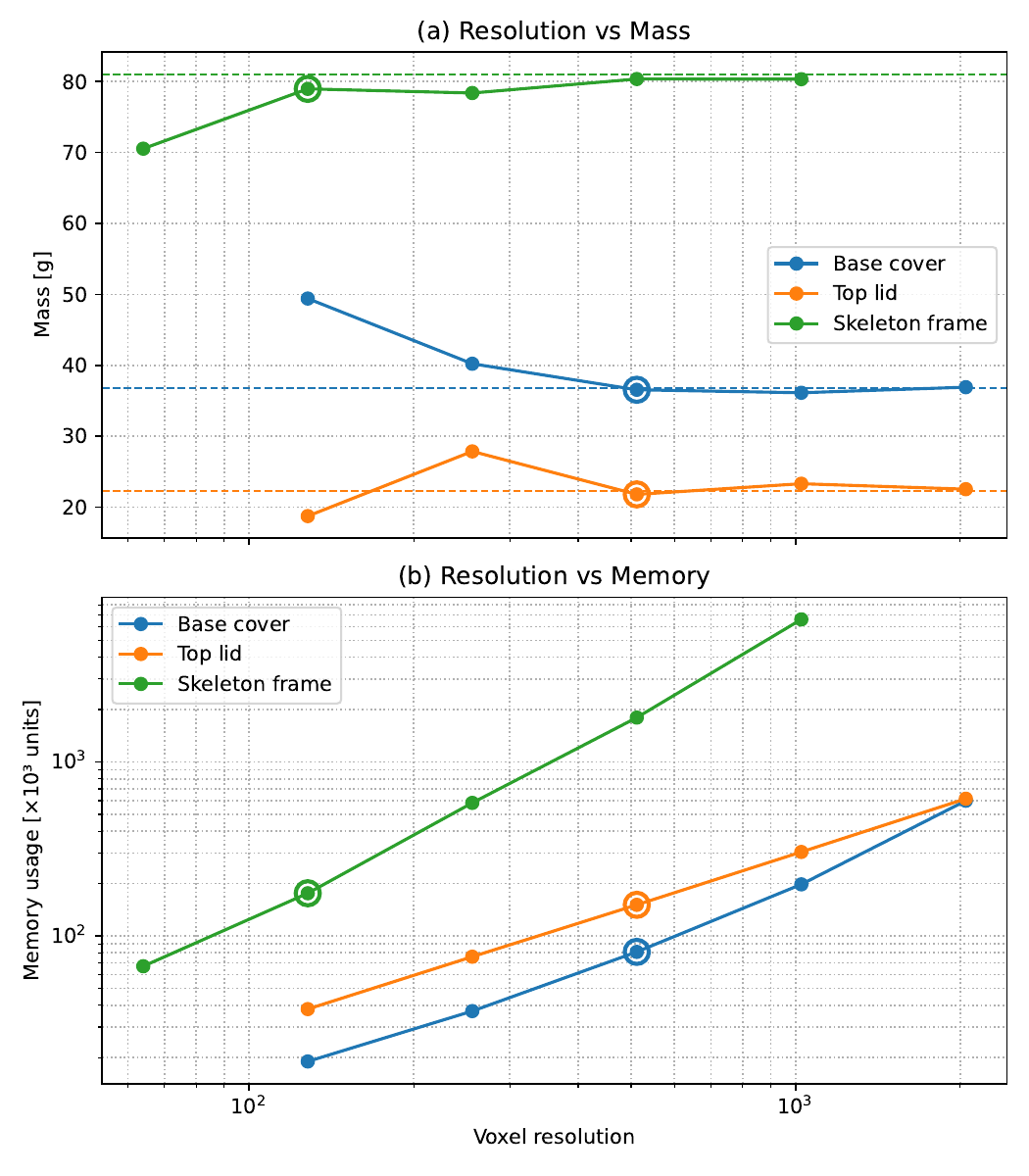}
\caption{Dependence of mass reproducibility and memory usage on voxel resolution for representative GRBAlpha structures. The adopted resolutions balance mass fidelity against computational cost.}
\label{fig:resolution_tradeoff}
\end{figure}

Knowing the exact weight and density of the real-world object additionally allows for an optimization on the lowest resolution for its most correct voxel approximation. Assuming that the voxelized object is uniform with respect to its own density and composition, the mass of each voxel can be coadded together and compared with the required weight of the real-world object. Different resolutions can then tested to find the stopping point where some optimal threshold is met. In practice, we adopted the minimum resolution for which the component mass agreed with the hardware-provided value to within approximately 5\%, while still keeping the resulting memory usage manageable. Figure~\ref{fig:resolution_tradeoff} shows this trade-off explicitly for the base cover, top plate, and skeleton frame.

\subsubsection{Export to MEGAlib geometry file}
For each brick with inclusive voxel bounds $(x_0,x_1,y_0,y_1,z_0,z_1)$ and physical voxel edge lengths $(\Delta_x,\Delta_y,\Delta_z)$ (in cm), we output a MEGAlib \texttt{BRIK}:
\[
\text{half-lengths } (h_x,h_y,h_z)=\frac{1}{2}\Big((x_1{-}x_0{+}1)\Delta_x,\,(y_1{-}y_0{+}1)\Delta_y,\,(z_1{-}z_0{+}1)\Delta_z\Big),
\]
\[
\text{center } (c_x,c_y,c_z)=\boldsymbol{\sigma}_\mathrm{cm} + \frac{1}{2}\Big((x_0{+}x_1{+}1)\Delta_x,\,(y_0{+}y_1{+}1)\Delta_y,\,(z_0{+}z_1{+}1)\Delta_z\Big),
\]
where $\boldsymbol{\sigma}_\mathrm{cm}$ is the voxelization shift converted to centimeters. Each brick becomes
\texttt{Volume <id>; <id>.Shape BRIK $h_x$ $h_y$ $h_z$; <id>.Position $c_x$ $c_y$ $c_z$; <id>.Material <mat>; <id>.Mother WorldVolume}.
We prepend a \texttt{WorldVolume} and a materials include (e.g., \texttt{Materials.geo}) as required by MEGAlib.

\subsection{Experimental Setup and Response Generation}
The Geant4 simulation framework is a toolkit library for particle transport codes based on Monte Carlo simulations, initially developed by CERN \cite{geant1,geant2,geant3}. All simulations in this work were based on Geant4 version 10.7.1, using the single Coulomb scattering (SS) model. To establish the framework to simulate the particle interaction with matter, we additionally used version 1.1 of the CADMesh library, which allows us to load triangular mesh-based CAD files into Geant4 \cite{cadmesh}. This combination of tools is particularly powerful, as it opens up many possibilities for future work and exploring the effects of different materials and treatments on particle interactions with matter.

The Medium Energy Gamma-ray Astronomy Library (MEGAlib) is a suite of software tools built on top of Geant4 designed to simulate, calibrate, and analyze data of hard X-ray and gamma-ray detectors, with a focus on Compton telescopes \citep{MEGAlib2006}. The software library contains all data analysis steps from simulation and measurements via calibration all the way to event and image reconstruction, with some high-level tools which allow to calculate response matrices and determine detector resolutions and sensitivites. The geometry and detector description tool, however, is limited to simple Geant4 shapes since integrating the CADMesh library codebase into MEGAlib is not trivial. Therefore, making a high-fidelity 3D approximation of any relatively complicated object requires simplification, and lot of manual work to describe each shape and composition. The procedure used has been already described in Section~\ref{sec:methods}.

\begin{figure}[h]
\centering
\includegraphics[width=0.74\linewidth]{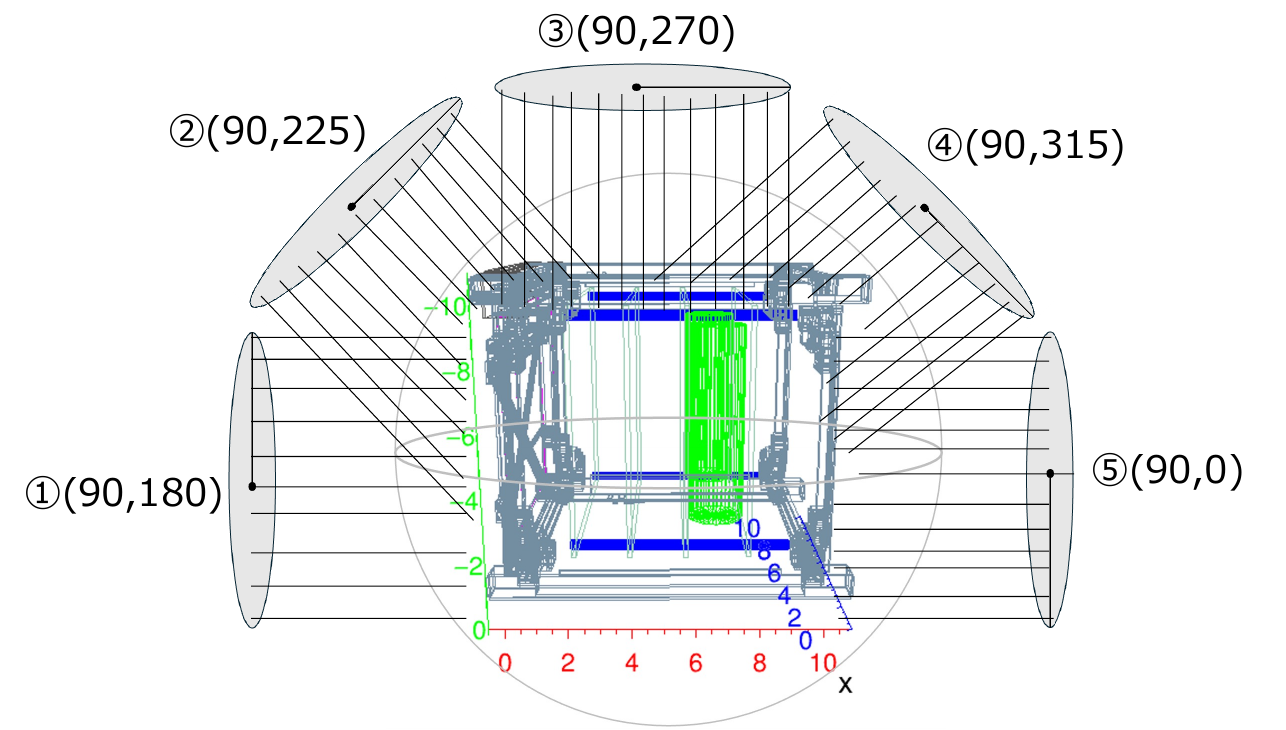}
\caption{The experimental setup for GRBAlpha, described in Table~\ref{tab:simconds}, of the different incident directions used to simulate the mono-energetic photons, and eventually to create the detector response.}
\label{fig:sim_angles}
\end{figure}

To construct detector response matrices, we followed a five-step workflow from the MEGAlib \texttt{cosima} outputs to an analysis-ready response product. Table~\ref{tab:simconds} summarizes the principal simulation settings adopted for GRBAlpha and VZLUSAT-2. First, mono-energetic simulations were performed for each incident direction and each incident energy. In these simulations, photons were irradiated as a parallel beam from a distant source using \texttt{FarFieldPointSource}, and the deposited energy in the CsI(Tl) scintillator was recorded for every triggered event. Repeating this procedure over the full incident-energy grid produced a set of output spectra that form the basis of the response calculation. An example of the experimental setup can be seen in Figure~\ref{fig:sim_angles}.

\begin{table}[h]
\centering
\small
\caption{Compact summary of the simulation conditions adopted for DRM generation.}
\label{tab:simconds}
\begin{tabular}{|l|c|c|c|c|c|}
\hline
Target & Energy range [keV] & Photons/energy & Angles & Disk radius [cm] & Channels \\ 
\hline
GRBAlpha & 0.1--1000 & $1.0\times10^5$ & 5 & 13.5 & 256 \\ 
VZLUSAT-2 & 0.1--2000 & $1.0\times10^5$ & 9 & 31 & 256 \\ 
\hline
\end{tabular}
\end{table}

Once a simulation is completed, the deposited photon energies are converted from physical energy units to detector channels using a linear calibration equation,
\[
\mathrm{ADC}=\frac{E_{\rm dep}-B}{A},
\]
where $A$ is the detector gain and $B$ is the offset. Since the measured spectra of the flight detectors have finite energy resolution, an additional Gaussian smearing is applied assuming an energy resolution of ${\rm FWHM}=6\%$, i.e.,
\[
\sigma(E)=\frac{0.06\,E}{2.355}\,,
\]
where 2.355 corresponds to $\rm{FWHM} = 2\sqrt{2\ln 2}\sigma \approx 2.355\sigma$.
This step yields simulated output spectra mapped to the same channel space used by the observational data products.

\begin{table}[h]
\centering
\small
\caption{Calibration coefficients used for converting deposited energy to detector channel.}
\label{tab:calib}
\begin{tabular}{|l|c|c|}
\hline
Detector & Gain $A$ & Offset $B$ \\ 
\hline
GRBAlpha (CsI) & 4.08 & -154.2 \\ 
VZLUSAT-2 Det0 & 9.062 & -362.917 \\ 
VZLUSAT-2 Det1 & 7.910 & -315.270 \\ 
\hline
\end{tabular}
\end{table}

It is also necessary to compute the effective-area spectrum from the simulated event statistics. If photons are uniformly generated on a disk of radius $R$, the total irradiation area is $S_{\rm total}=\pi R^2$. For a simulation with $N_{\rm total}$ generated photons and $N_{\rm trigger}$ events that deposit energy in the CsI, the effective area is described by,
\[
S_{\rm eff}=S_{\rm total}\frac{N_{\rm trigger}}{N_{\rm total}}.
\]
In this way, the channel histogram at each incident energy is converted from counts to effective area in cm$^2$, yielding $S_{\rm eff}(E_i,{\rm ch}_j)$. This quantity establishes the baseline as the validation metric used between the original Geant4 and MEGAlib simulations.

With a successful validation, the two-dimensional DRM can then be constructed by stacking the effective-area spectra over all mono-energetic simulations,
\[
{\rm DRM}(E_i,{\rm ch}_j)=S_{\rm eff}(E_i,{\rm ch}_j).
\]
The resulting matrix represents the probabilistic mapping between incident photon energy and detector output channel, including the effects of shielding, scattering, and detector resolution. 

Finally, the DRM was converted into an OGIP-compliant FITS response file so that it can be used directly in standard high-energy analysis software. We implemented this conversion in Python using \texttt{astropy.io.fits}, writing the response into the \texttt{MATRIX} extension together with the corresponding \texttt{ENERG\_LO}, \texttt{ENERG\_HI}, \texttt{F\_CHAN}, \texttt{N\_CHAN}, and \texttt{MATRIX} columns, and adding the channel-boundary information in the \texttt{EBOUNDS} extension. This provides a reproducible path from CAD-based spacecraft models to simulation-derived response files that can be distributed independently of the original Geant4 or MEGAlib environment.

\section{Geant4 and MEGAlib validation}
We first applied the above workflow to GRBAlpha as a benchmark case for validating the MEGAlib-based response construction. Starting from an initial geometry model, we progressively refined the material definitions, the Pb-shield configuration, and the relative placement of the battery, PCB boards, support rods, and detector-adjacent structures. These changes were particularly important for low-energy photons, for which small differences in shielding thickness or geometry produce large changes in the number of photons that can reach the CsI scintillator. After optimization, the agreement between the MEGAlib and Geant4 responses improved substantially across the main energy band, demonstrating that the detector response can be reproduced reliably when the dominant shielding structures are modeled with sufficient fidelity. 

\begin{figure}[h]
\centering
\begin{minipage}[t]{0.49\linewidth}
    \centering
    \includegraphics[width=\linewidth]{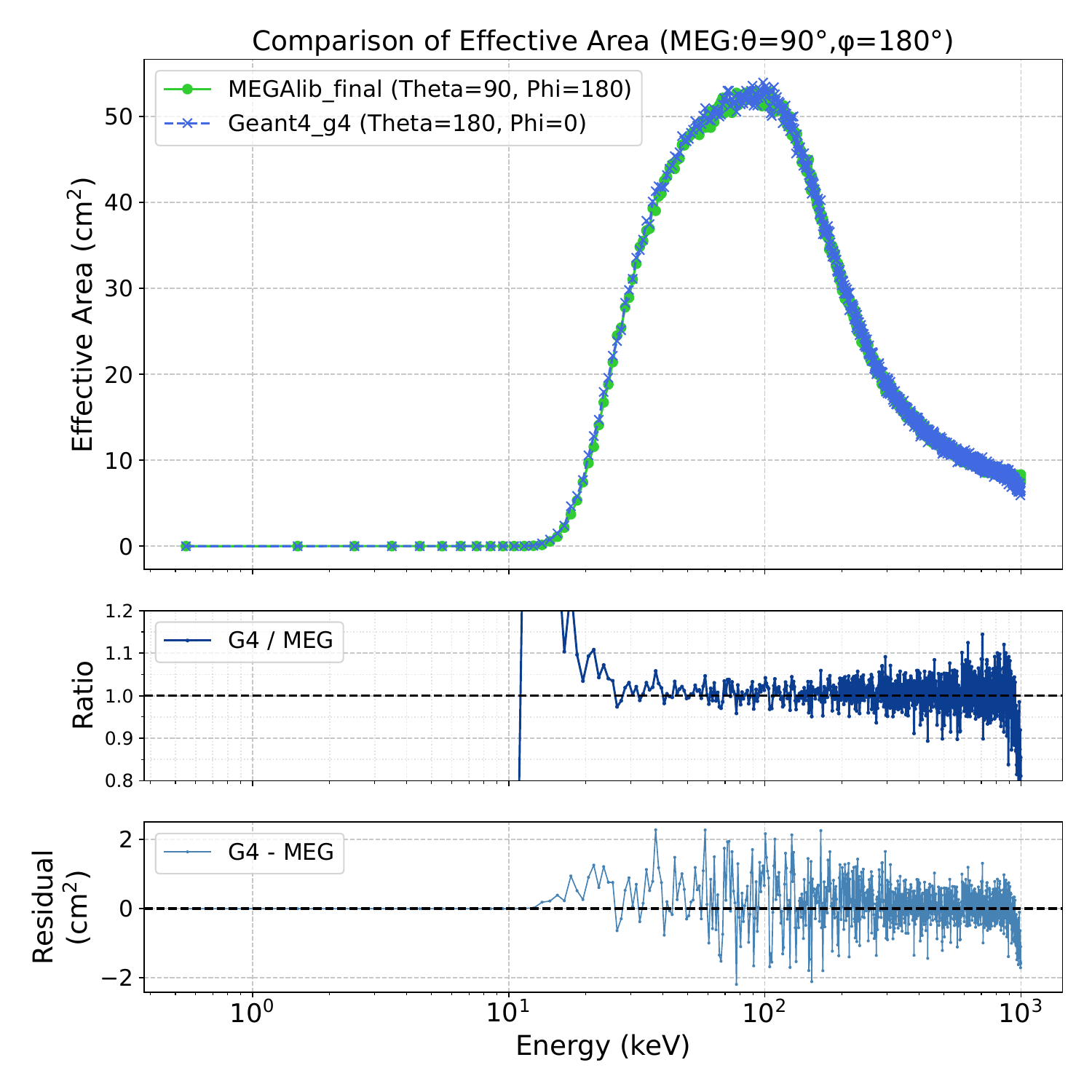}
\end{minipage}
\hfill
\begin{minipage}[t]{0.49\linewidth}
    \centering
    \includegraphics[width=\linewidth]{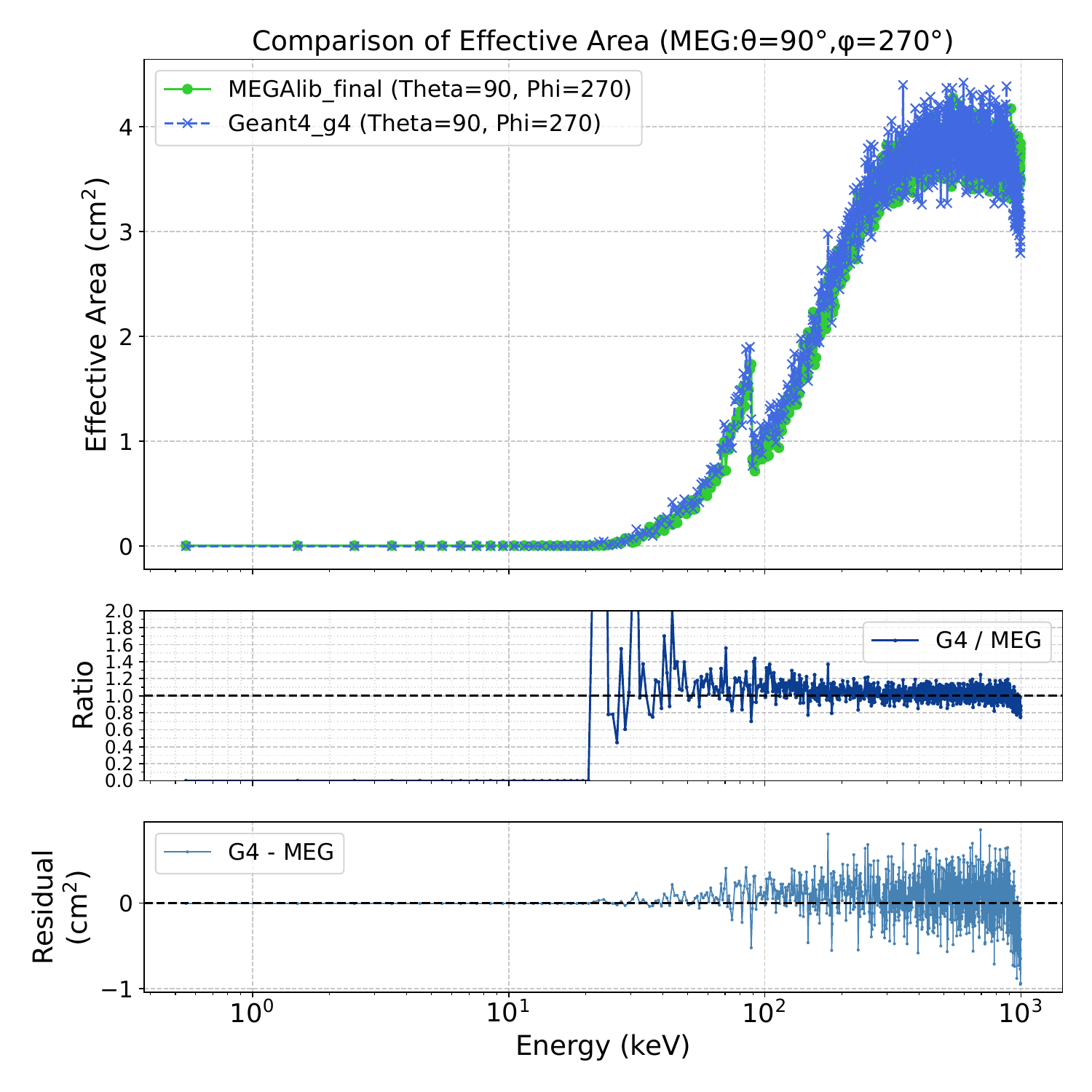}
\end{minipage}
\caption{Representative GRBAlpha effective-area comparisons between MEGAlib and Geant4 for two incident directions. Left: face-on detector impact, $(\theta,\varphi)=(90^\circ,180^\circ)$. Right: lead shield-facing impact, $(\theta,\varphi)=(90^\circ,270^\circ)$. The representative face-on ratio-distribution analysis gives a residual systematic difference of about 3\%, while both panels show good overall agreement between the MEGAlib and Geant4 effective-area curves.}
\label{fig:grbalpha_compare_both}
\end{figure}

For GRBAlpha, we compared the MEGAlib-derived response with an existing Geant4 reference DRM for five representative irradiation directions. In this case, the distribution of the Geant4/MEGAlib effective-area ratio is centered close to unity, with no significant systematic bias indicating that one framework consistently overestimates or underestimates the response. Figure~\ref{fig:grbalpha_compare_both} shows the comparison of the Effective Area between the original Geant4 and the MEGAlib simulations for a face-on direct-angle impact to the detector (left), and an angle directly impacting the lead shielding above the detector (right), corresponding to angles $(\theta,\varphi)=(90^\circ,180^\circ)$ and $(\theta,\varphi)=(90^\circ,270^\circ)$.

For the representative face-on case, using the fitted width of the ratio distribution $\sim3.5\%$ and subtracting the representative statistical contribution $\sim1.7\%$ in quadrature yields a residual systematic difference of $\sim3\%$, supporting the validity of the MEGAlib simulation setup. This representative $\sim3\%$ value should be distinguished from the broader curve-level agreement, which is assessed from the effective-area comparisons themselves.

\begin{figure}[h]
\centering
\includegraphics[width=0.54\linewidth]{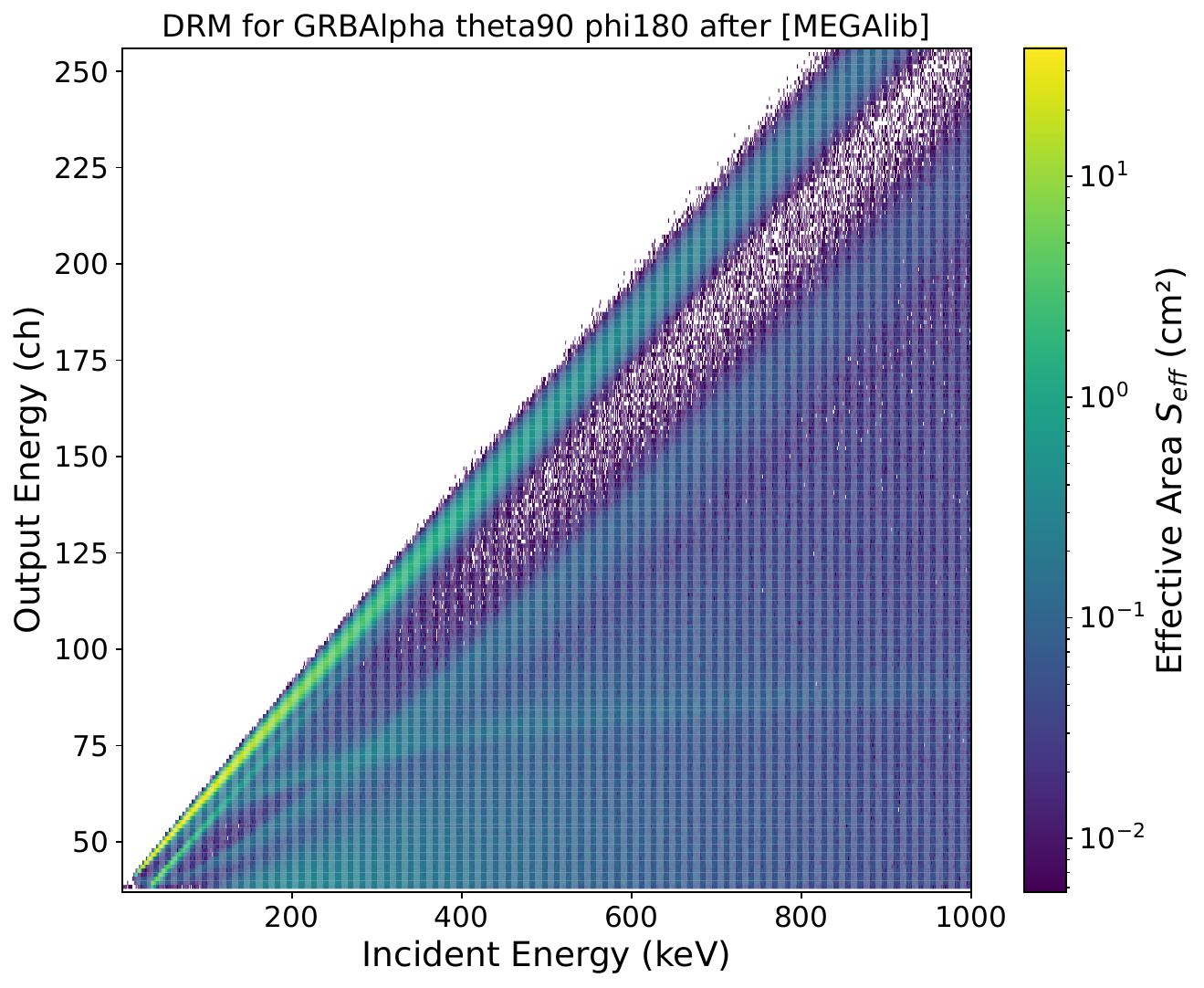}
\caption{Example of the two-dimensional response matrix generated using a range of mono-energetic simulations for the face-on detector orientation corresponding to $(\theta,\phi)=(90^\circ,180^\circ)$.}
\label{fig:drm}
\end{figure}

The resulting two-dimensional response matrix for the face-on detector impact $(\theta,\varphi)=(90^\circ,180^\circ)$ over a range of energies is also shown in Figure~\ref{fig:drm}.

\subsection{Validation of simple and voxel geometry}
With the MEGAlib simulation framework validated, the same procedure was applied to VZLUSAT-2, whose 3U structure is considerably more complex than that of GRBAlpha and contains many additional shielding and scattering components. For this satellite,  DRMs were constructed for nine representative incident directions and for both detector units. The results show a clear and systematic dependence of the effective area on the incident direction and on which detector face is exposed more directly to the incoming photons. In particular, differences between front-side and back-side incidence are strongest at low energies, where attenuation by solar panels, enclosure structures, and internal components is largest, while the differences decrease toward higher energies as the photons become more penetrating.

\begin{figure}[h]
\centering
\begin{minipage}[t]{0.49\linewidth}
    \centering
    \includegraphics[width=\linewidth]{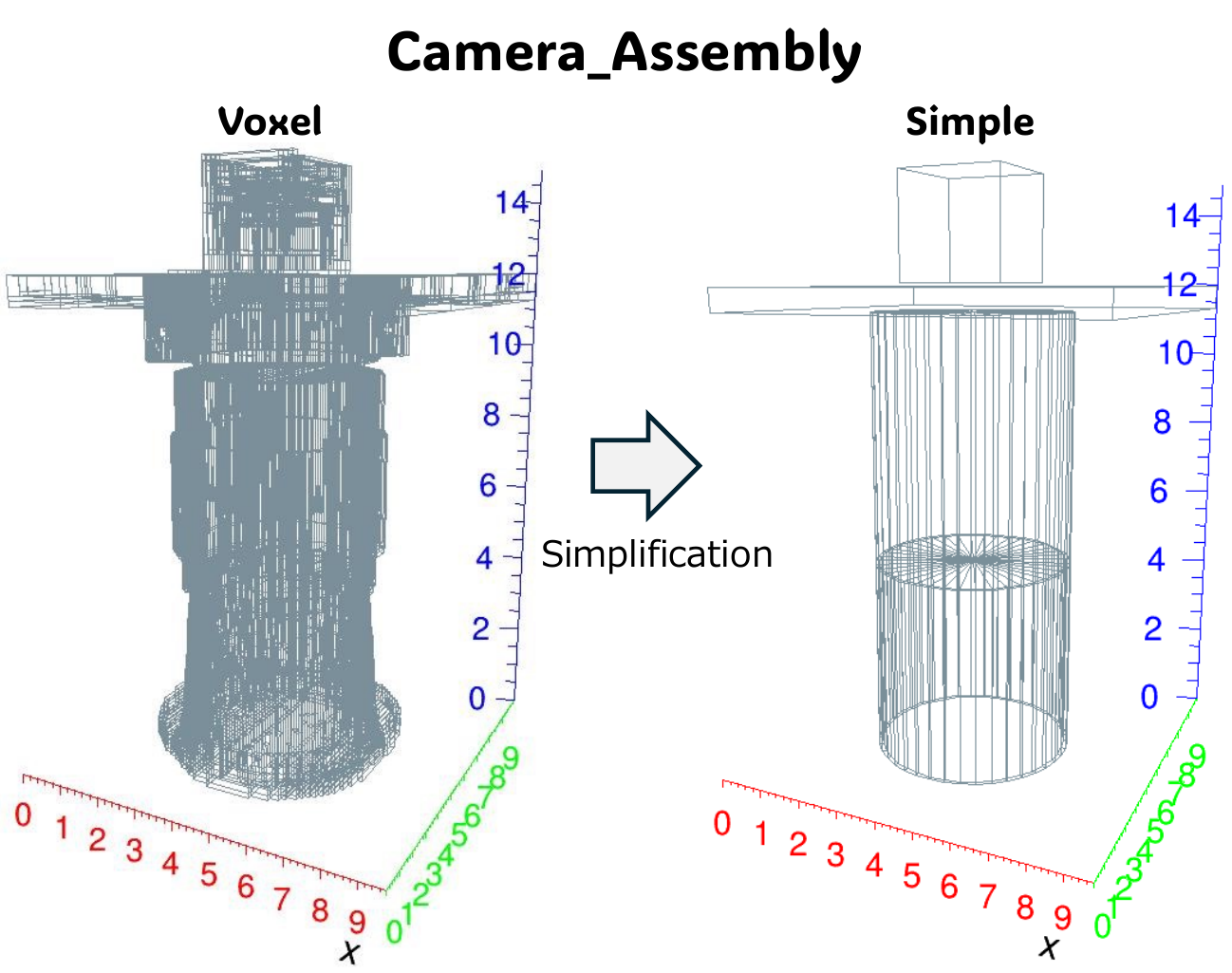}
\end{minipage}
\hfill
\begin{minipage}[t]{0.49\linewidth}
    \centering
    \includegraphics[width=\linewidth]{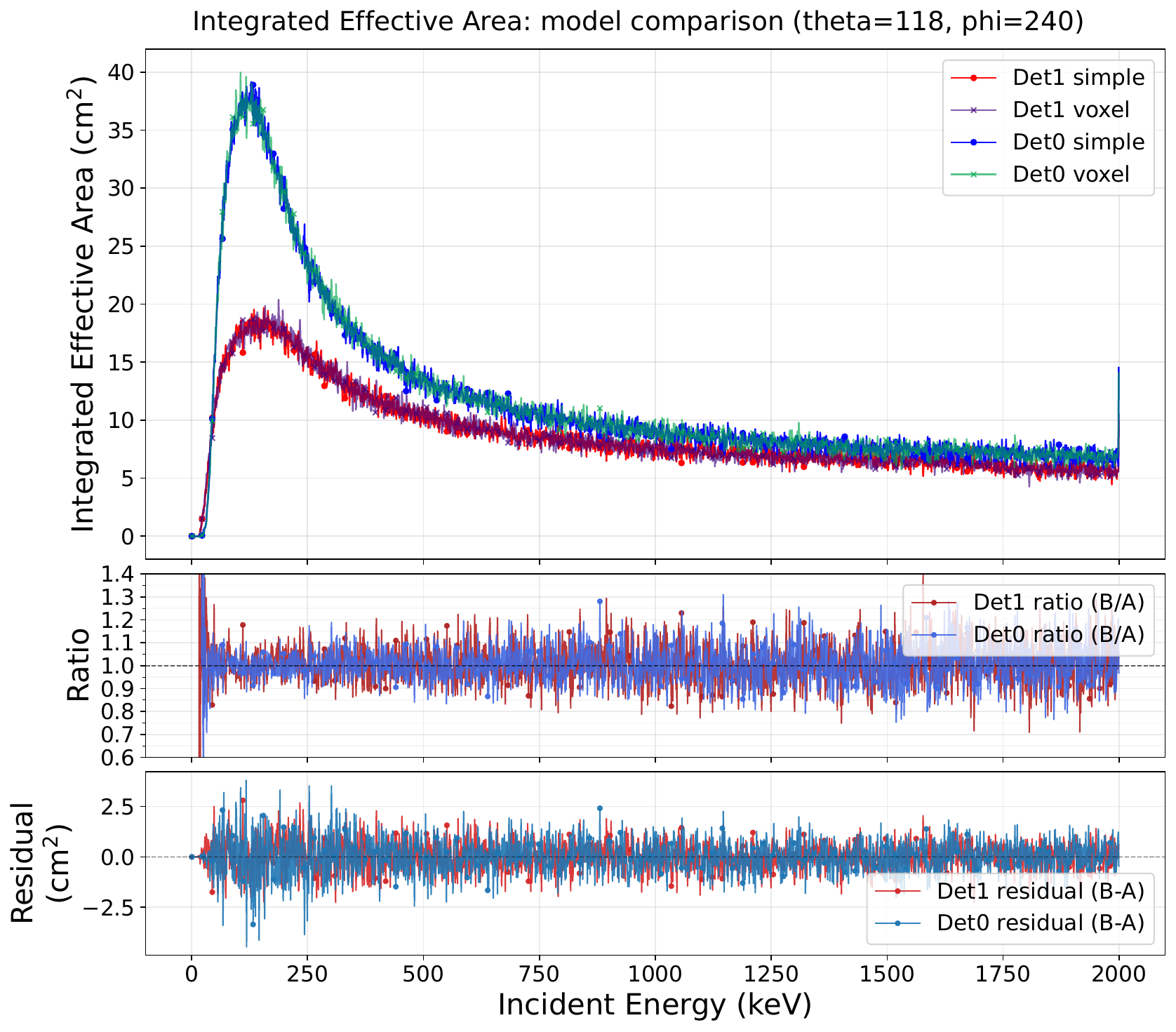}
\end{minipage}
\caption{Left: example of voxelized camera assembly geometry vs a simplified model. Right: representative comparisons between the simple and optimized-voxel VZLUSAT-2 response models corresponding to $(\theta,\phi)=(118^\circ,240^\circ)$. In the main energy range, the simplified geometry reproduces the voxel-optimized response closely while substantially reducing computation time.}
\label{fig:vzlu_compare}
\end{figure}

The full, multi-angle calculations with optimized voxel geometries are computationally expensive, so a simplified geometry description for VZLUSAT-2 was also evaluated as an alternative. Comparison with the voxel-optimized model showed that the simplified model reproduces the main response behavior well in the principal energy band while reducing the computational burden significantly. This confirms that a hybrid strategy is effective: high-fidelity voxelization should be concentrated on detector-adjacent structures and other components that dominate the shielding, whereas more distant or geometrically simple components can be approximated with primitive shapes at much lower cost. A compact comparison of the three VZLUSAT-2 geometry tiers is given in Table~\ref{tab:vzlu_models}, while Figure~\ref{fig:vzlu_compare} (left) shows an example of the geometry simplification for the camera assembly on VZLUSAT-2, as well as a representative simple-vs-voxel response comparison from a simulation corresponding to $(\theta,\phi)=(118^\circ,240^\circ)$ (right).

\begin{table}[h]
\centering
\small
\caption{Compact summary of the VZLUSAT-2 geometry tiers used in this study.}
\label{tab:vzlu_models}
\begin{tabular}{|l|c|c|l|}
\hline
Model & Total memory [kB] & Time/angle [h] & Role in this work \\ 
\hline
High-res. voxel & 3237 & $\gtrsim 8000$ & Highest-fidelity reference; impractical for multi-angle DRM \\ 
Optimized voxel & 620 & 84--100 & Accuracy benchmark for detector-adjacent structures \\ 
Simple geometry & 350 & 23--28 & Production model for multi-angle DRM construction \\ 
\hline
\end{tabular}
\end{table}

For the representative direction $(\theta,\varphi)=(118^\circ,240^\circ)$, the ratio distributions indicate systematic differences of about $4.5\%$ for Det0 and $5.8\%$ for Det1, while the simplified model reduces the computation time by roughly a factor of four, consistent with the total simulation-cost summary in Table~\ref{tab:vzlu_models}. Taken together, these results show that the proposed geometry-conversion workflow can deliver accurate DRMs for both simple and complex CubeSat structures, while retaining the flexibility needed for large multi-angle simulation campaigns.

\section{Conclusions and Future Activities} 
In this work, a practical MEGAlib-based workflow was developed for constructing detector response matrices of CubeSat gamma-ray detectors from detailed spacecraft mass models. The core of the method is a geometry-conversion procedure in which CAD-derived components are voxelized, compressed into rectangular bricks, and exported to a MEGAlib-compatible description while preserving the shielding and scattering structures most relevant to the detector response. This makes it possible to generate multi-angle DRMs even for satellites whose geometries are too complex to represent manually with primitive solids alone.

Using GRBAlpha as a benchmark, the MEGAlib-derived response can reproduce an existing Geant4 reference response well, even with complex, voxelized geometries. By refining key shielding structures and material definitions, we achieved good agreement between the Geant4 and MEGAlib responses in the main energy band. Moreover, for the representative GRBAlpha face-on angle $(90^\circ,180^\circ)$ case, the ratio-distribution analysis gives a residual systematic difference of about $3\%$. Applying the same workflow to VZLUSAT-2 demonstrated that the method also scales to a more complex 3U platform with two detectors and strong direction dependence. For the representative angle $(118^\circ,240^\circ)$, a simplified geometry reproduces the optimized-voxel response within approximately $6\%$ while reducing the simulation time by roughly a factor of four, which is important for constructing response sets over many incident directions.

These results establish a response-generation framework that is sufficiently accurate for practical use and flexible enough to support future CAMELOT analyses. The next steps are to increase the angular sampling of the DRMs, optimize photon statistics according to incident direction and energy, and extend the same workflow to additional CAMELOT satellites. For VZLUSAT-2 in particular, the DRMs prepared for the two candidate detector orientations can be combined with observational Det0/Det1 ratios and spacecraft attitude information to further constrain the in-orbit detector configuration and ultimately improve GRB arrival-direction estimation in both spacecraft and sky coordinates.

\acknowledgments 
NW, JR, MD, and FM thank the support by the Czech Science Foundation (GAČR) project No. 24-11487J. MD is a Brno Ph.D. Talent Scholarship Holder---Funded by the Brno City Municipality. This work is also supported by the National Science and Technology Council (NSTC) of Taiwan under grants 113-2923-M-007-004-MY3. This work is also supported by JSPS KAKENHI Grant Numbers JP21KK0051.

\bibliography{main} 
\bibliographystyle{spiebib} 

\end{document}